\documentstyle [12pt] {article}
\begin{document}
\setcounter{page}{1}
\def\theequation{\arabic{section}.\arabic{equation}}
\def\theequation{\thesection.\arabic{equation}}
\setcounter{section}{0}

\title{On the $D^+_s \to \pi^+ \pi^+ \pi^-$-nonresonant decay in the
 effective quark model with chiral symmetry}

\author{A. N. Ivanov\thanks{E--mail: ivanov@kph.tuwien.ac.at, Tel.:
+43--1--58801--5598, Fax: +43--1--5864203}~{\footnotesize $^{^{\dag}}$} and
N.I. Troitskaya\thanks{Permanent Address:
State Technical University, Department of Theoretical
Physics, 195251 St. Petersburg, Russian Federation}}

\date{}

\maketitle

\begin{center}
{\it Institut f\"ur Kernphysik, Technische Universit\"at Wien, \\
Wiedner Hauptstr. 8-10, A-1040 Vienna, Austria}
\end{center}

\vskip1.0truecm
\begin{center}
\begin{abstract}
The partial widths of the decays $D^+_s \to \phi\,\pi^+$ and $D^+_s \to
\pi^+ \pi^+ \pi^-$-nonresonant are computed within the effective quark
model with chiral symmetry involving Heavy quark effective theory (HQET)
and Chiral perturbation theory at the quark level (CHPT)$_q$ with linear
realization of chiral $U(3)\times U(3)$ symmetry. It is shown that for the
explanation of the experimental probability of the $D^+_s \to \pi^+ \pi^+
\pi^-$--nonresonant decay one does not need to use unnaturally heavy light
current quarks as has been suggested in Ref.[1].
\end{abstract}
\end{center}
\vspace{0.2in}

\begin{center}
PACS number(s): 13.30.Ce, 12.39.Ki, 14.20.Lq, 14.20.Mr
\end{center}

\vskip1.0truecm

\newpage

\section{Introduction}
\setcounter{equation}{0}

The problem of the $D^+_s \to (\pi^+ \pi^+ \pi^-)_{\rm NR}$--nonresonant
(NR) decay has been recently discussed in Ref.[1]. According to the
conclusion suggested in Ref.[1] the theoretical explanation of the
experimental probability of this decay can be reached via the adoption of
unnaturally heavy light
current quark masses $\bar{m}=\frac{1}{2}\,(m_{0 u} + m_{0 d}) = 22\div 38\;
{\rm MeV}$ [1] instead of the widely accepted value $\bar{m}=\frac{1}{2}\,
(m_{0 u} + m_{0 d}) =5.5\;{\rm MeV}$ [2].

The application of the effective quark model with chiral symmetry involving
Heavy quark effective theory (HQET) [3-5] supplemented by Chiral
perturbation theory at the quark level (CHPT)$_q$  [6-8] with linear
realization of chiral $U(3)\times U(3)$ symmetry, to the calculation  of
chiral corrections to the mass spectra and leptonic constants of charmed
mesons [9], and to the form factors of semileptonic $D$--meson decays [10]
has shown that the widely accepted values of the current quark masses
$m_{0u}=4\;{\rm MeV}$, $m_{0d}=7\;{\rm MeV}$ and $m_{0s}=135\;{\rm MeV}$
[2,6-8] describe well the experimental data.

This paper is to apply the effective quark model with chiral $U(3)\times
U(3)$ symmetry incorporating HQET and (CHPT)$_q$ to the computation of the
partial width and the probability of the $D^+_s \to (\pi^+ \pi^+
\pi^-)_{\rm NR}$ decay. Since experimentally the probability of this decay
has been defined compared
the probability of the $D^+_s\to \phi \pi^+$ decay [11], i.e.
\begin{eqnarray}\label{label1.1}
\frac{\Gamma(D^+_s\to (\pi^+ \pi^+ \pi^-)_{\rm NR})}{\Gamma(D^+_s\to \phi
\pi^+)} = 0.29 \pm 0.09 \pm 0.06,
\end{eqnarray}
we also compute the partial width of the $D^+_s\to \phi \pi^+$ decay.

The effective low--energy Lagrangian responsible for non--leptonic decays of
the $D^+_s$--meson reads [12]
\begin{eqnarray}\label{label1.2}
{\cal L}_{\rm eff}(x) = -\frac{G_F}{\sqrt{2}}\,V^*_{c s}\,V_{u d}\,
C_1\,[\bar{s}\,\gamma^{\mu}(1-\gamma^5)\,c(x)]\,
[\bar{u}\,\gamma_{\mu}(1-\gamma^5)\,d(x)],
\end{eqnarray}
where $G_F=1.166\times 10^{-5}\;{\rm GeV}^{-2}$ is the Fermi weak constant,
$V^*_{c s}$ and $V_{u d}$ are the elements of the CKM--mixing matrix and
$C_1\simeq 1.3$ is the Wilson coefficient caused by the strong quark--gluon
interactions at scales $p > \mu$, where $\mu$ is a normalization scale. In
(CHPT)$_q$ we should identify $\mu$ with the scale of spontaneous breaking
of chiral symmetry (SB$\chi$S) $\Lambda_{\chi}=0.94\;{\rm GeV}$ [7], i.e.
$\mu = \Lambda_{\chi}=0.94\;{\rm GeV}$ (see also [6,8]). The contribution
of strong interactions at scales $p \le \mu=\Lambda_{\chi}$ is described by
(CHPT)$_q$ in terms of constituent quark loop diagrams, where the  momenta
of virtual quarks are restricted from above by the SB$\chi$S scale
$\Lambda_{\chi}$ [6--10,13--16].

The effective quark model incorporating HQET and (CHPT)$_q$ resembles that
suggested by Bardeen and Hill [17] that is also based on the
Nambu--Jona--Lasinio model. There is only distinction that in the
Bardeen--Hill model heavy--light mesons are considered like partners of
light mesons, whereas in our effective quark model, heavy mesons are
external states with respect to the light ones. This distinction influences
only the redefinition of phenomenological parameters that are introduced in
the model. Nevertheless, all results obtained within our effective quark
model should be  valid too in the Bardeen--Hill model.

\section{The  $D^+_s\to \phi \pi^+$ decay}
\setcounter{equation}{0}

The amplitude of the non--leptonic decay $D^+_s\to \phi \pi^+$ can be
defined as follows
\begin{eqnarray}\label{label2.1}
M(D^+_s\to \phi \pi^+) = <\pi^+(q) \phi(Q)|{\cal L}_{\rm eff}(0)|D^+_s(p)>.
\end{eqnarray}
In (CHPT)$_q$ at the tree--meson approximation the computation of the
matrix element of the non--leptonic decay $D^+_s\to \phi \pi^+$
Eq.(\ref{label2.1}) agrees with the vacuum insertion approximation leading
to a factorized
amplitude [7]. This gives
\begin{eqnarray}\label{label2.2}
&&M(D^+_s\to \phi \pi^+) = <\pi^+(q) \phi(Q)|{\cal L}_{\rm
eff}(0)|D^+_s(p)>=\nonumber\\
&&= -\frac{G_F}{\sqrt{2}}\,V^*_{c s}\,V_{u d}\,C_1\,<\pi^+(q)|
[\bar{u}\,\gamma^{\mu}(1-\gamma^5)\,d(x)]|0>\nonumber\\
&& \times
<\phi(Q)|[\bar{s}\,\gamma_{\mu}(1-\gamma^5)\,c(x)]|D^+_s(p)>=\nonumber\\
&&=-\,i\,G_F\,V^*_{c s}\,V_{u d}\,C_1\,F_{\pi}\,q^{\mu}\,
<\phi(Q)|[\bar{s}\,\gamma_{\mu}(1-\gamma^5)\,c(x)]|D^+_s(p)>.
\end{eqnarray}
The computation of the matrix element
$<\pi^+(q)|[\bar{u}\,\gamma^{\mu}(1-\gamma^5)\,d(x)]|0>$  has been
carried out in Ref.[7]
\begin{eqnarray}\label{label2.3}
<\pi^+(q)|[\bar{u}\,\gamma^{\mu}(1-\gamma^5)\,d(x)]|0> = i\,
\sqrt{2}\,F_{\pi}\,q^{\mu},
\end{eqnarray}
where $F_{\pi} \simeq 92\;{\rm MeV}$ is a leptonic constant of the
$\pi^+$--meson [11].

The matrix element
$<\phi(Q)|[\bar{s}\,\gamma_{\mu}(1-\gamma^5)\,c(x)]|D^+_s(p)>$ of the
$D^+_s\to \phi$ transition can be expressed in
 terms of the form factors of the semileptonic
$D^+_s\to \phi\,{\ell}\,\nu_{\ell}$ decay [10,16]
\begin{eqnarray}\label{label2.4}
&&<\phi(Q)|[\bar{s}\,\gamma_{\mu}(1-\gamma^5)\,c(x)]|D^+_s(p)> =\nonumber\\
&&=i a_1(q^2)\,e^*_{\mu}(Q) -i\,a_2(q^2)\,(e^* (Q) \cdot p)\,(p + Q)_{\mu} -
 i\,a_3(q^2)\,(e^* (Q) \cdot p)\,(p - Q)_{\mu} \nonumber\\
&&- 2\, b(q^2)\, \varepsilon_{\mu\nu\alpha\beta} \,e^{*\nu}(Q)\,
p^{\alpha}Q^{\beta}, \;(\varepsilon^{0123} = 1),
\end{eqnarray}
where $e^*_{\mu}(Q)$ is the polarization 4--vector of the $\phi$--meson. As
a result the amplitude of the $D^+_s\to \phi \pi^+$ decay reads
\begin{eqnarray}\label{label2.5}
&&M(D^+_s\to \phi\pi^+) = \nonumber\\
&&=G_F\,V^*_{c s}\,V_{u d}\,C_1\,F_{\pi}\,[ a_1(M^2_{\pi}) - (M^2_{D^+_s} -
M^2_{\phi})\,a_2(M^2_{\pi}) - M^2_{\pi}a_3(M^2_{\pi})]\,q\cdot e^* (Q),
\end{eqnarray}
where we have set $p^2 = M^2_{D^+_s}$, $Q^2 = M^2_{\phi}$ and  $q^2 =
M^2_{\pi}$. Setting $M_{\pi}=0$ we reduce the r.h.s. of Eq.(\ref{label2.5})
to the form
\begin{eqnarray}\label{label2.6}
&&M(D^+_s \to \phi \pi^+) = \nonumber\\
&&=G_F\,V^*_{c s}\,V_{u d}\,C_1\,F_{\pi}\,[ a_1(0) - (M^2_{D^+_s} -
M^2_{\phi})\,a_2(0)]\,q\cdot e^* (Q).
\end{eqnarray}
In HQET supplemented by (CHPT)$_q$ the form factors $a_i(0)$ ($i=1,2$)
have been computed in Ref.[16] and read
\begin{eqnarray}\label{label2.7}
a_1(0) = 1.47\;{\rm GeV}\quad,\quad a_2(0) = 0.21\;{\rm GeV}^{-1}.
\end{eqnarray}
The partial width of the $D^+_s \to \phi \pi^+ $ decay is given by
\begin{eqnarray}\label{label2.8}
&&\Gamma(D^+_s \to \phi \pi^+) =|C_1|^2 |G_F V^*_{c s} V_{u d}|^2
F^2_{\pi}\Big[a_1(0) - (M^2_{D^+_s} - M^2_{\phi})\,a_2(0)\Big]^2
\frac{|\vec{q}\,|^3}{8\pi M^2_{D^+_s}}=\nonumber\\
&&=|C_1|^2 |G_F V^*_{c s} V_{u d}|^2 F^2_{\pi}\,
\frac{\Big[a_1(0) -
(M^2_{D^+_s} - M^2_{\phi})\,a_2(0)\Big]^2}{64 \pi}
\frac{M^4_{D^+_s}}{M^3_{\phi}}\Bigg(1 -
\frac{M^2_{\phi}}{M^2_{D^+_s}}\Bigg)^3 = \nonumber\\
&&=|C_1|^2 \times 0.36\times 10^{11}\;{\rm s}^{-1},
\end{eqnarray}
where $|\vec{q}\,| = (M^2_{D^+_s} - M^2_{\phi})/2M_{\phi}$ is a relative
momentum of a $\phi$--meson and a massless $\pi^+$--meson. The numerical
value has been obtained at $|V_{c s}|=1.01$, $|V_{u d}|=0.975$,
$M_{D^+_s}=1.97\;{\rm GeV}$, $M_{\phi}=1.02\;{\rm GeV}$ [11]. We have
dropped the dependence on the $\pi$--meson mass and set $M_{\pi}=0$.

When using the partial width Eq.(\ref{label2.8}) we can estimate
$\tau_{D^+_s}$ the time life of the $D^+_s$--meson:
\begin{eqnarray}\label{label2.9}
\tau_{D^+_s} &=& {\displaystyle \frac{B(D^+_s \to \phi \pi^+)}{|C_1|^2
\times 0.36\times 10^{11}}} = {\displaystyle
\frac{(3.6\pm 0.9)\times 10^{-2}}{|C_1|^2 \times 0.36\times 10^{11}}}
\nonumber\\
&=& (0.59\pm 0.15)\times 10^{-12}\,{\rm s},
\end{eqnarray}
where we have applied the experimental value $B(D^+_s \to \phi \pi^+) =
(3.6\pm 0.9)\,\%$ [11]. Our estimate $\tau_{D^+_s}=(0.59\pm 0.15)\times
10^{-12}\,{\rm s}$ agrees well with the experimental value
$\tau_{D^+_s}=(0.467\pm 0.017)\times 10^{-12}\,{\rm s}$.

\section{The  $D^+_s\to (\pi^+ \pi^+ \pi^-)_{\rm NR}$ decay}
\setcounter{equation}{0}

The amplitude of the $D^+_s\to (\pi^+ \pi^+ \pi^-)_{\rm NR}$ decay is  defined
\begin{eqnarray}\label{label3.1}
&&M(D^+_s\to (\pi^+\pi^+\pi^-)_{\rm NR}) = <(\pi^+(q_+) \pi^+(p_+)
\pi^-(p_-))_{\rm NR}|{\cal L}_{\rm eff}(0)|D^+_s(p)> = \nonumber\\
&&=-\frac{G_F}{\sqrt{2}} V^*_{c s} V_{u d} C_1 <(\pi^+(q_+) \pi^+(p_+)
\pi^-(p_-))_{\rm NR}|\bar{u}(0)\,\gamma_{\mu}(1 -
\gamma^5)\,d(0)|0>\nonumber\\
&&\times <0|\bar{s}(0)\,\gamma^{\mu}(1 - \gamma^5)\,c(0)|D^+_s(p)>.
\end{eqnarray}
The matrix element $<0|\bar{s}(0)\,\gamma^{\mu}(1 -
\gamma^5)\,c(0)|D^+_s(p)>$ is determined by the $D^+_s$ leptonic constant
$F_{D^+_s}$:
\begin{eqnarray}\label{label3.2}
<0|\bar{s}(0)\,\gamma^{\mu}(1 - \gamma^5)\,c(0)|D^+_s(p)> =
-\,i\,\sqrt{2}\,F_{D^+_s}\,p^{\mu},
\end{eqnarray}
where $F_{D^+_s} = 1.17\,F_{D^+} = 1.41\,F_{\pi}$ [9,13].
Thus the amplitude of the $D^+_s\to (\pi^+\pi^+\pi^-)_{\rm NR}$ decay reads
\begin{eqnarray}\label{label3.3}
&&M(D^+_s\to (\pi^+\pi^+\pi^-)_{\rm NR}) = i\,G_F\,V^*_{c s}\,V_{u
d}\,C_1\,F_{D^+_s} \nonumber\\
&& \times p_{\mu}\,<(\pi^+(q_+) \pi^+(p_+) \pi^-(p_-))_{\rm NR}|\bar{u}(0)
\,\gamma^{\mu}(1 - \gamma^5)\,d(0)|0>.
\end{eqnarray}
In (CHPT)$_q$ the matrix element $<(\pi^+(q_+) \pi^+(p_+) \pi^-(p_-))_{\rm
NR}|\bar{u}(0)\gamma_{\mu}(1 - \gamma^5) d(0)|0>$ is defined by the
box--constituent quark diagrams [8] and can be described by the momentum
integral
\begin{eqnarray}\label{label3.4}
&&<(\pi^+(q_+) \pi^+(p_+) \pi^-(p_-))_{\rm NR}|\bar{u}(0)\,\gamma^{\mu}
(1 - \gamma^5)\,d(0)|0>= i\,\frac{N g^3_{\pi}}{16\,\pi^2}\times\nonumber\\
&&\int \frac{d^4k}{\pi^2 i}\,{\rm tr}\Bigg\{\gamma^{\mu}\gamma^5
\frac{1}{m - \hat{k} + \hat{p}}\gamma^5
\frac{1}{m - \hat{k} + \hat{p} - \hat{p}_+}\gamma^5
 \frac{1}{m - \hat{k} + \hat{q}_+}\gamma^5\frac{1}{m - \hat{k}}\Bigg\}
\nonumber\\
&& + (q_+ \leftrightarrow p_+) + \ldots ,
\end{eqnarray}
where the ellipses denote the contribution of the matrix element of the
vector current $\bar{u}(0)\,\gamma^{\mu}\,d(0)$ not contributing to the
$D^+_s\to (\pi^+\pi^+\pi^-)_{\rm NR}$ decay amplitude, then $m=0.33\;{\rm
GeV}$ is the constituent light quark mass calculated in the chiral limit
[6--8], and $g_{\pi}=\sqrt{2}\,m/F_{\pi}$, the coupling constant of
quark--pion interaction, satisfies the constraint $N g^2_{\pi}/8\pi^2=1$
[13]. The former becomes rather obvoius, if one would take into account
that
$F_{\pi}=O(\sqrt{\displaystyle N})$ at $N\to \infty$ and make a change
$F_{\pi}\to F_{\pi}\,\sqrt{\displaystyle N/3}$, where in the r.h.s. of this
relation $F_{\pi}=92\;{\rm MeV}$. Then, one finds substituting the
numerical data that $Ng^2_{\pi}/8\pi^2 = 3m^2/4\pi^2 F^2_{\pi} \simeq 1$.
We should empasize that we have ignored the contibution of the
$\pi^+$--meson pole contribution to the r.h.s. of Eq.(\ref{label3.4}). As
it is shown in Appendix the contribution of the $\pi^+$--meson pole is of
order $O(1/M^4_{D^+_s})$ compared with the contribution of the momentum
integral left in Eq.(\ref{label3.4}).

Since in (CHPT)$_q$ [8] virtual momenta of constiuent quark loops are
restricted by the SB$\chi$S scale, i.e. $k\le \Lambda_{\chi}$, and HQET
assumes that $M_{D^+_s} \simeq M_c$, so in a heavy--quark mass limit we
following the Appelquist--Carazzone theorem [22] reduce the r.h.s. of
Eq.(\ref{label3.4}) to the form
\begin{eqnarray}\label{label3.5}
&&<(\pi^+(q_+) \pi^+(p_+) \pi^-(p_-))_{\rm NR}|\bar{u}(0)\,
\gamma^{\mu}(1 - \gamma^5)\,d(0)|0>= \nonumber\\
&&= - \frac{i}{\sqrt{2}F_{\pi}}\,\frac{m^2}{M^2_{D^+_s}}\int
\frac{d^4k}{\pi^2 i}\,{\rm tr}\Bigg\{\gamma^{\mu}\gamma^5
\frac{1}{m - \hat{k} + \hat{q}_+}\gamma^5 \frac{1}{m - \hat{k}}\Bigg\} +
 (q_+ \leftrightarrow p_+),
\end{eqnarray}
where we have used the relations $g_{\pi}=\sqrt{2}\,m/F_{\pi}$, $N
g^2_{\pi}/8\pi^2=1$ and $p^2 = M^2_{D^+_s}$. The calculation of the
integral over $k$ gives one
\begin{eqnarray}\label{label3.6}
\int \frac{d^4k}{\pi^2 i}\,{\rm tr}\Bigg\{\gamma^{\mu}\gamma^5
\frac{1}{m - \hat{k} + \hat{q}_+}\gamma^5 \frac{1}{m - \hat{k}}\Bigg\}=
 4\,m\,q^{\mu}_+\,\frac{16\pi^2}{Ng^2_{\pi}}\,g^2_{\pi}I_2(m).
\end{eqnarray}
In the r.h.s. of Eq.(\ref{label3.6}) we encounter a logarithmically
divergent integral which is well defined in (CHPT)$_q$  by the
compositeness condition $g^2_{\pi}I_2(m) = 1$ [6--9] (see also
Refs.[25,26]), where we have used the definition of $I_2(m)$ [6--9]
\begin{eqnarray}\label{label3.7}
I_2(m)=\frac{N}{16\pi^2}\int \frac{d^4k}{\pi^2i}\frac{1}{(m^2 - k^2)^2},
\end{eqnarray}
and by applying the relation $Ng^2_{\pi}/8\pi^2 =1$ [13]. The compositeness
condition $g^2_{\pi}I_2(m) = 1$ realizes correct kinetic terms of
low--lying mesons, $\bar{q}q$--collective excitations, in effective chiral
Lagrangians derived in effective quark models based on the
Nambu--Jona--Lasinio approach [6--9,17,23,24].

Formally, the calculation of the momentum integrals representing one--loop
constituent quark diagrams should be carried out keeping only the divergent
parts and dropping the contributions of the parts finite in the infinite
limit of the cut--off. Such a prescription realizes a naive description of
confinement of quarks. Indeed, dropping the finite parts of quark diagrams
we remove the imaginary parts of them and suppress by this the appearance
of quarks in the intermediate states of low--energy hadron interactions.
This naive description of confinement has turned out to be rather useful
for the derivation of effective Chiral Lagrangians [6--9,23,24]. As has
been shown in Ref. [21] this prescription can be justified in the
multicolour QCD ($N\to \infty$) with a linearly rising interquark
confinement potential. Thereby, within such a naive approach to the quark
confinement mechanism one can bridge  quantatively the quark and the hadron
level of the description of strong low--energy interactions of hadrons. The
cut--off $\Lambda_{\chi}=0.94\,{\rm GeV}$ and the constituent quark mass
$m=0.33\,{\rm GeV}$ can be considered in such an approach as input
parameters and fixed at one--loop approximation via the experimental values
of the $\rho\pi\pi$ coupling constant $g_{\rho}$ and the leptonic pion
constant $F_{\pi}$ [6]. Thus, we should accentuate that in the effective
quark models based on the Nambu--Jona--Lasinio approach quark diagrams lose
the meaning of quantum field theory objects and only display how quark
flavours can be transferred form an initial hadron state to a final hadron
state in hadron--hadron low--energy transitions. The coupling constants of
such transitions desribed in terms of divergent parts of quark diagrams and
depending of the cut--off $\Lambda_{\chi}=0.94\,{\rm GeV}$ and the
constituent quark mass  $m=0.33\,{\rm GeV}$ can be expressed in terms of
effective phenomenological coupling constants of low--energy hadron
interactions given by effective chiral Lagrangians [6--9,17,23,24]. Hence,
such a description of strong low--energy interaction of hadrons can be
valued as good established as the effective chiral Lagrangian approach. As
has been shown in Refs.[6--10,14--16] (CHPT)$_q$ allows to develop
consistent expansions in powers of both current quark masses and external
momenta of interacting hadrons, i.e. chiral perturbation theory at both
tree--meson and one--loop meson level, in good agreement with experimental
data.

For the matrix element $<(\pi^+(q_+) \pi^+(p_+) \pi^-(p_-))_{\rm
NR}|\bar{u}(0)\,\gamma_{\mu}(1 - \gamma^5)\,d(0)|0>$ we obtain the
following
\begin{eqnarray}\label{label3.8}
<(\pi^+(q_+) \pi^+(p_+) \pi^-(p_-))_{\rm NR}|\bar{u}(0)\,\gamma^{\mu}(1 -
\gamma^5)\,d(0)|0>= - i \frac{4\sqrt{2}}{F_{\pi}}\,\frac{m^2}{M^2_{D^+_s}}
\,(p_+ + q_+)^{\mu}.
\end{eqnarray}
This gives one
\begin{eqnarray}\label{label3.9}
p_{\mu}<(\pi^+(q_+) \pi^+(p_+) \pi^-(p_-))_{\rm NR}|\bar{u}(0)\,
\gamma^{\mu}(1 - \gamma^5)\,d(0)|0>= - i
\frac{2\sqrt{2}}{F_{\pi}}\,\frac{m^2}{M^2_{D^+_s}}\,(M^2_{D^+_s} + q^2),
\end{eqnarray}
where we have set $q^2_+ = p^2_+ = p^2_- = 0$ and $q^2 = (p_+ + q_+)^2$.

Thus, in our approach the r.h.s. of Eq.(\ref{label3.9}) is not proportional
to the sum of current quark masses
\begin{eqnarray}\label{label3.10}
&&p_{\mu}<(\pi^+(q_+) \pi^+(p_+) \pi^-(p_-))_{\rm NR}|\bar{u}(0)\,
\gamma^{\mu}(1 - \gamma^5)\,d(0)|0>= \nonumber\\
&&=- (m_{0u} + m_{0d})<(\pi^+(q_+) \pi^+(p_+) \pi^-(p_-))_{\rm NR}
|\bar{u}(0)\,\gamma^5\,d(0)|0>,
\end{eqnarray}
that can be expected when assuming the validity of the application of the
equations of motion for the free current quark fields, i.e. $\bar{u}(0)
\hat{p}(1 - \gamma^5) d(0) = - (m_{0u} + m_{0d}) \bar{u}(0) \gamma^5 d(0)$.
This entails a vanishing of the matrix element $p_{\mu}<(\pi^+(q_+)
\pi^+(p_+) \pi^-(p_-))_{\rm NR}|\bar{u}(0) \gamma^{\mu}(1 - \gamma^5)
d(0)|0>$ in the chiral limit. Let us show now that at $p^2 =
M^2_{D^+_s}\not= M^2_{\pi}$ the matrix element $p_{\mu}<(\pi^+(q_+)
\pi^+(p_+) \pi^-(p_-))_{\rm NR}|
\bar{u}(0)\,\gamma^{\mu}(1 - \gamma^5)\,d(0)|0>$ does not vanish in the
chiral limit.

Let us consider a more general matrix element  $<\pi^+(q_+) \pi^+(p_+)
\pi^-(p_-)|\bar{u}(0)\,\gamma^{\mu}(1 - \gamma^5)\,d(0)|0>$  and represent
it in terms of invariant amplitudes
\begin{eqnarray}\label{label3.11}
&&- i <\pi^+(q_+) \pi^+(p_+) \pi^-(p_-)|\bar{u}(0)\,
\gamma^{\mu}(1 - \gamma^5)\,d(0)|0> = \nonumber\\
&&\hspace{1in} = F_1 \, (p_+ + q_+)^{\mu} + F_2 \, p^{\mu} + i F_3 \,
\varepsilon^{\mu\nu\alpha\beta}p_{\nu}p_{+\alpha}q_{+\beta},
\end{eqnarray}
where $F_i\,(i=1,2,3)$ are invariant amplitudes free of kinematical
sigularities. Multiplying both sides of Eq.(\ref{label3.11}) by a momentum
$p_{\mu}$ we arrive at the expression
\begin{eqnarray}\label{label3.12}
&&- i p_{\mu}<\pi^+(q_+) \pi^+(p_+) \pi^-(p_-)|\bar{u}(0)\,
\gamma^{\mu}(1 - \gamma^5)\,d(0)|0> = \nonumber\\
&&\hspace{1in} = \frac{1}{2}F_1 \,(p^2 - p^2_-  + q^2) + F_2 \, p^2.
\end{eqnarray}
Imposing then the Adler condition, i.e. demanding a vanishing of the
amplitude at $p_-\to 0$, we obtain a relation between invariant amplitudes
$F_2 = - F_1$, which ensues the equation
\begin{eqnarray}\label{label3.13}
- i \lim_{p_-\to 0}p_{\mu}<\pi^+(q_+) \pi^+(p_+)
\pi^-(p_-)|\bar{u}(0)\,\gamma^{\mu}(1 - \gamma^5)\,d(0)|0> =
\frac{1}{2}(F_1 + F_2) p^2 = 0,
\end{eqnarray}
where we have used the relation $q^2 \to p^2$ valid in the limit $p_-\to 0$
due to a conservation of 4--momentum $p = q_+ + p_+ + p_-$. Using the
relation $F_2 = - F_1$ we get
\begin{eqnarray}\label{label3.14}
- i p_{\mu}<\pi^+(q_+) \pi^+(p_+) \pi^-(p_-)|\bar{u}(0)\,
\gamma^{\mu}(1 - \gamma^5)\,d(0)|0> = \frac{1}{2} \,F_1\,
(q^2 - p^2 - p^2_-).
\end{eqnarray}
The correctness of this expression can be verified by matching
Eq.(\ref{label3.14}) with the amplitude of the elastic
$\pi^+\pi^+$--scattering ($\pi^+$ + $\pi^+$ $\to$  $\pi^+$ + $\pi^+$).
Setting $p^2 = p^2_- = M^2_{\pi}$ and denoting $q^2 = (p_+ + q_+)^2 = s$ we
reduce the r.h.s. of Eq.(\ref{label3.14}) to the form
\begin{eqnarray}\label{label3.15}
- i p_{\mu}<\pi^+(q_+) \pi^+(p_+) \pi^-(p_-)|\bar{u}(0)\,
\gamma^{\mu}(1 - \gamma^5)\,d(0)|0> = \frac{1}{2}\,F_1\,(s - 2\,M^2_{\pi}),
\end{eqnarray}
which up to the common factor agrees well with the amplitude
${\cal M}(\pi^+ + \pi^+ \to \pi^+ + \pi^+) = (2\,M^2_{\pi}-s)/F^2_{\pi}$
 given by Weinberg.

In the chiral limit $p^2_+ = q^2_+ = p^2_- = M^2_{\pi} \to 0$ we find
\begin{eqnarray}\label{label3.16}
- ip_{\mu}<\pi^+(q_+) \pi^+(p_+) \pi^-(p_-)|\bar{u}(0)\,
\gamma^{\mu}(1 - \gamma^5)\,d(0)|0> = \frac{1}{2}F_1 s.
\end{eqnarray}
It is correct, since nothing suppresses the elastic $\pi^+\pi^+$--scattering
for massless $\pi^+$--mesons. This confirms the validity of
 Eq.(\ref{label3.9}) and non--vanishing behaviour of the matrix element
$p_{\mu}<(\pi^+(q_+) \pi^+(p_+) \pi^-(p_-))_{\rm NR}|\bar{u}(0)\,
\gamma^{\mu}(1 - \gamma^5)\,d(0)|0>$ in the chiral limit at $p^2 =
M^2_{D^+_s}\not= M^2_{\pi}$.

Substituting Eq.(\ref{label3.9}) in Eq.(\ref{label3.3}) we get the amplitude
 of the $D^+_s\to (\pi^+\pi^+\pi^-)_{\rm NR}$ decay amplitude
\begin{eqnarray}\label{label3.17}
M(D^+_s\to (\pi^+\pi^+\pi^-)_{\rm NR}) = C_1\,2\sqrt{2}m^2\,G_F\,V^*_{c s}
\,V_{u d}\,\frac{F_{D^+_s}}{F_{\pi}}\,\frac{M^2_{D^+_s} + q^2}{M^2_{D^+_s}}.
\end{eqnarray}
The partial width of the $D^+_s\to (\pi^+\pi^+\pi^-)_{\rm NR}$ decay computed
at $M_{\pi}=0$ is given by
\begin{eqnarray}\label{label3.18}
\Gamma(D^+_s\to (\pi^+\pi^+\pi^-)_{\rm NR}) &=& =|C_1|^2\,|G_F\,V^*_{c s}
\,V_{u d}|^2\,\frac{11
m^4}{384\pi^3}\,\frac{F^2_{D^+_s}}{F^2_{\pi}}\,M_{D^+_s}=\nonumber\\
&&=|C_1|^2\times 0.87\times 10^{10}\;{\rm s}^{-1},
\end{eqnarray}
where the numerical value has been obtained at $m=0.33\;{\rm GeV}$,
$|V_{c s}|=1.01$, $|V_{u d}|=0.975$ and $M_{D^+_s}=1.97\;{\rm GeV}$.

Matching this value with the value of the partial width of the
$D^+_s\to \phi \pi^+$ decay we arrive at the ratio
\begin{eqnarray}\label{label3.19}
\frac{\Gamma(D^+_s\to (\pi^+ \pi^+ \pi^-)_{\rm NR})}{\Gamma(D^+_s\to \phi
\pi^+)} = 0.24\pm \Delta,
\end{eqnarray}
which agrees well with the experimental data Eq.(\ref{label1.1}).

\section{Conclusion}

The application of the effective quark model with chiral $U(3)\times U(3)$
symmetry incorporating HQET and (CHPT)$_q$ to the computation of the
partial widths of the $D^+_s\to \phi \pi^+$ and $D^+_s\to (\pi^+ \pi^+
\pi^-)_{\rm NR}$ decays has shown the agreement of the theoretical values
with the experimental data. We have found $\Gamma(D^+_s\to (\pi^+ \pi^+
\pi^-)_{\rm NR})/\Gamma(D^+_s\to \phi \pi^+) = 0.24\pm 0.12$, which agrees
well with the experimental value $\Gamma(D^+_s\to (\pi^+ \pi^+ \pi^-)_{\rm
NR})/\Gamma(D^+_s\to \phi \pi^+) = 0.29\pm 0.09 \pm 0.06$.

Thus, one can conclude that such an effective quark model describes
reasonably well a low--energy dynamics of heavy--light meson interactions,
and one does not need to include unnaturally heavy light current quarks in
order to explain the experimental data on the $D^+_s\to (\pi^+ \pi^+
\pi^-)_{\rm NR}$ decay [1].

Using the experimental probability of the  $D^+_s\to \phi \pi^+$ decay,
i.e. $Br.(D^+_s\to \phi \pi^+) = (3.6\pm 0.9)\,\%$, we have estimated the
time life of the $D^*_s$--meson: $\tau_{D^+_s} = (0.59\pm 0.15)\times
10^{-12}\;{\rm s}$. This theoretical value agrees well with the
experimental one: $(\tau_{D^+_s})_{\exp} = (0.467\pm 0.017)\times
10^{-12}\;{\rm s}$ [11].

As has been discussed in Ref.[14] the theoretical uncertainty of our
effective quark model is about 50$\%$. However, for all cases of the
application of this model, the resultant agreement between theoretical and
experimental data turns out as usually much better.

Recall that our results have been obtained for the factorized amplitudes.
The step beyond the factorization approximation is to take into account
one--meson loop contributions. In (CHPT)$_q$ the procedure of the
computation of one--meson loop corrections has been considered in Ref.[6]
(Ivanov).  We are planning to investigate such contributions in our
forthcoming publications.

\section{Appendix. The $\pi^+$--meson pole contribution}
\setcounter{equation}{0}

In (CHPT)$_q$ the contribution of the $\pi^+$--meson pole to the matrix
element Eq.(\ref{label3.4}) is given by [8]
\begin{eqnarray}\label{label5.1}
&&i\,<(\pi^+(q_+) \pi^+(p_+) \pi^-(p_-))_{\rm NR}|\bar{u}(0)\,
\gamma^{\mu}(1 - \gamma^5)\,d(0)|0>_{\pi^+-{\rm pole}}= \nonumber\\
&&=\frac{N g_{\pi}}{16\pi^2}\int \frac{d^4k}{\pi^2 i} {\rm
tr}\Bigg\{\gamma^{\mu}\gamma^5 \frac{1}{m - \hat{k} + \hat{p}}\gamma^5
\frac{1}{m - \hat{k} }\Bigg\}\frac{1}{p^2-M^2_{\pi}}
\Bigg(-\frac{N g^4_{\pi}}{16\pi^2}\Bigg)
\int \frac{d^4k}{\pi^2 i}\nonumber\\
&&{\rm tr}\Bigg\{\gamma^5 \frac{1}{m - \hat{k} + \hat{p}}\gamma^5
\frac{1}{m - \hat{k} + \hat{p} - \hat{p}_+}\gamma^5
\frac{1}{m - \hat{k} + \hat{q}_+}\gamma^5\frac{1}{m - \hat{k}}\Bigg\} +
(q_+ \leftrightarrow p_+) .
\end{eqnarray}
Taking into account that $M_{D^+_s}\gg \Lambda_{\chi}\ge k$ we reduce
the r.h.s. of Eq.(\ref{label5.1}) to the form
\begin{eqnarray}\label{label5.2}
&&i\,<(\pi^+(q_+) \pi^+(p_+) \pi^-(p_-))_{\rm NR}|\bar{u}(0)\,
\gamma^{\mu}(1 - \gamma^5)\,d(0)|0>_{\pi^+-{\rm pole}}=
 \frac{1}{2\sqrt{2}}\,\frac{m}{F_{\pi}}\,\frac{1}{M^4_{D^+_s}}\nonumber\\
&&\int \frac{d^4k}{\pi^2 i} {\rm tr}\Bigg\{\gamma^{\mu}\hat{p}
 \frac{1}{m - \hat{k} }\Bigg\}\frac{1}{p^2-M^2_{\pi}}
\int \frac{d^4k}{\pi^2 i}
{\rm tr}\Bigg\{\gamma^5 \frac{1}{m - \hat{k} + \hat{q}_+}\gamma^5
\frac{1}{m - \hat{k}}\Bigg\}\nonumber\\
&&+ (q_+ \leftrightarrow p_+),
\end{eqnarray}
where we have used the relations $g_{\pi}=\sqrt{2}\,m/F_{\pi}$, $N
g^2_{\pi}/8\pi^2=1$ and $p^2 = M^2_{D^+_s}$.

The momentum integrals entering the r.h.s. of Eq.(\ref{label5.2}) equal [6--9]
\begin{eqnarray}\label{label5.3}
&&\int \frac{d^4k}{\pi^2 i} {\rm tr}\Bigg\{\gamma^{\mu}\hat{p}
 \frac{1}{m - \hat{k} }\Bigg\} = 4\,m^2\,\bar{v}\,p^{\mu},\nonumber\\
&&\int \frac{d^4k}{\pi^2 i}
{\rm tr}\Bigg\{\gamma^5 \frac{1}{m - \hat{k} + \hat{q}_+}\gamma^5
\frac{1}{m - \hat{k}}\Bigg\}= 4\,m\,\bar{v},
\end{eqnarray}
where $\bar{v}= - <0|\bar{q}q|0>/F^2_{\pi} = 1.92\,{\rm GeV}$ [6--9].
This gives the following contribution of the $\pi^+$--pole
\begin{eqnarray}\label{label5.4}
&&i\,<(\pi^+(q_+) \pi^+(p_+) \pi^-(p_-))_{\rm NR}|\bar{u}(0)\,
\gamma^{\mu}(1 - \gamma^5)\,d(0)|0>_{\pi^+-{\rm pole}}=\nonumber\\
&&= 8\sqrt{2}\,\frac{m^4}{F_{\pi}}\,\frac{\bar{v}^2}{M^6_{D^+_s}}\,p^{\mu}.
\end{eqnarray}
The contribution of the $\pi^+$--meson pole to the amplitude
 Eq.(\ref{label3.3}) is given by
\begin{eqnarray}\label{label5.5}
M(D^+_s\to (\pi^+\pi^+\pi^-)_{\rm NR})_{\pi^+-{\rm pole}} =
 C_1\,2\sqrt{2}\,m^2\,G_F\,V^*_{c s}\,V_{u d}\,
\frac{F_{D^+_s}}{F_{\pi}}\,\frac{2 m^2\bar{v}^2}{M^4_{D^+_s}}.
\end{eqnarray}
Thus, we have shown that the contribution of the $\pi^+$--meson pole is of
order $O(1/M^4_{D^+_s})$ compared with the contribution of the momentum
integral Eq.(\ref{label3.5}) and can be ignored, correspondingly.


\newpage
\begin{thebibliography}{9}
\bibitem{[1]}
Hoang N. L., Nguyen A. V. and Pham X. Y.,
{\it Phys. Lett. B}, 357 (1995) 177 and references therein.
\bibitem{[2]}
Gasser J. and H. Leutwyler H.,
{\it Phys. Rep.}, 87 (1982) 77.
\bibitem{[3]}
Eichten E. and Feinberg F. L.,
{\it Phys. Rev. D}, 23 (1981) 2724;
Eichten E.,
{\it Nucl. Phys. B}, 4 (Proc.Suppl.) (1988) 70;
Voloshin M. B. and Shifman M. A.,
{\it Sov. J. Nucl. Phys.}, 45 (1987) 292;
\bibitem{[4]}
Politzer H. D. and Wise M.,
{\it Phys. Lett. B}, 206 (1988); ibid. {\it B}, 208 (1988) 504.
\bibitem{[5]}
Georgi H.,
{\it Phys. Lett. B}, 240 (1990) 447.
\bibitem{[6]}
Ivanov A. N., Nagy M. and Troitskaya N. I.,
{\it Int. J. Mod. Phys. A}, 7 (1992)  7305;
Ivanov A. N.,
{\it Int. J. Mod. Phys. A}, 8 (1993) 853.
\bibitem{[7]}
Ivanov A. N.,
{\it Phys. Lett. B}, 275 (1992) 450;
Ivanov A. N., Troitskaya N. I. and Nagy M.,
{\it Int. J. Mod. Phys. A}, 8 (1993) 2027, 3425.
\bibitem{[8]}
Ivanov A. N. and Troitskaya N. I.,
{\it Nuovo Cim. A}, 108 (1995) 555.
\bibitem{[9]}
Ivanov A. N. and Troitskaya N. I.,
{\it Phys. Lett. B}, 342 (1995) 323.
\bibitem{[10]}
Hussain F., Ivanov A. N. and Troitskaya N. I.,
{\it Phys. Lett. B}, 348 (1995) 609 ; ibid. {\it B}, 369 (1996) 351.
\bibitem{[11]}
Partical Data Group,
{\it Phys. Rev. D}, 54(1996) 1, Part1.
\bibitem{[12]}
Lee B. W. and Gaillard M. K.,
{\it Phys. Rev. Lett.}, 33 (1974) 108;
Altarelli G., Curci G., Martinelli G.and Petrarca S.,
{\it Nucl. Phys. B}, 187 (1981) 461;
Buras A., G$\grave{{\rm e}}$rard J.- M. and  R\"uckl R.,
{\it Nucl. Phys. B}, 268 (1986) 16;
Bauez M., Stech B.and Wizbel M.,
{\it Z. Phys. C}, 34 (1987) 103.
\bibitem{[13]}
Ivanov A. N., Troitskaya N. I. and Nagy M.,
{\it Phys. Lett. B}, 339 (1994) 167;
Hussain F., Ivanov A. N. and Troitskaya N. I.,
{\it Phys. Lett. B}, 329 (1994) 98; ibid. {\it B}, 334 (1994) E450;
\bibitem{[14]}
Ivanov A. N. and Troitskaya N. I.,
{\it  Phys. Lett. B}, 345 (1995) 175; ibid.{\it B}, 390 (1997) 341.
\bibitem{[15]}
Ivanov A. N. and Troitskaya N. I.,
{\it Nuovo Cim. A}, 110 (1997) 65.
\bibitem{[16]}
Ivanov A. N. and Troitskaya N. I.,
{\it Phys. Lett. B}, 394 (1997) 195.
\bibitem{[17]}
Bardeen W. A. and Hill C. T.,
{\it Phys. Rev.D}, 49 (1994) 409.
\bibitem{[18]}
Nambu  Y. and Jona--Lasinio G.,
{\it Phys. Rev.}, 122 (1961) 345; ibid. 124 (1961) 246.
\bibitem{[19]}
Eguchi T,
{\it Phys. Rev. D}, 14 (1976) 2755;
Kikkawa K.,
{\it Progr. Theor. Phys.}, 56 (1976) 947;
Kleinert H.,
{\it Proc. of Int. Summer School of Subnuclear Physics}, Erice 1976, Ed.
A.Zichichi, p.289.
\bibitem{[20]}
Klint S., Lutz M., Vogl V. and Weise W.,
{\it Nucl. Phys. A}, 516 (1990) 429, 469  and references therein.
\bibitem{[21]}
Ivanov A. N., Troitskaya N. I., Faber M., Schaler M. and Nagy M.,
{\it Nuovo Cim. A}, 107 (1994) 1667;
{\it Phys. Lett. B}, 336 (1994) 555.
\bibitem{[22]}
Appelquist  T. and Carazzone J.,
{\it  Phys. Rev. D}, 11 (1975) 2856.
\bibitem{[23]}
Dhar A.and Wadia S. R.,
{\it Phys. Rev. Lett.},  52 (1984) 959;
Dhar A., Shankar R. and Wadia S. R.,
{\it Phys. Rev. D}, 31 (1985) 3256;
Ebert D. and Reinhardt H.,
{\it Nucl. Phys. B}, 271 (1986) 188;
Wakamatsu M.,
{\it Ann. Phys.}, 193 (1989) 287;
Ruiz Arriola E.,
{\it Phys. Lett. B}, 264 (1991) 178.
\bibitem{[24]}
Bijnens J., Bruno C. and de Rafael E.,
{\it Nucl. Phys. B}, 390 (1993) 501  and references therein;
Bijnens J., de Rafael E. and Zheng H.,
{\it Z. Phys. C}, 62 (1994) 437 and references therein.
\bibitem{[25]}
Hayashi K., Hirayama M., Muta T., Seto N. and Shirafuji T.,
{\it Fortschritte der Physik}, 15 (1967) 625.
\bibitem{[26]}
Efimov G. V. and Ivanov M. A.,
in {\it Quark confinement model of hadrons},
Institute of Physics Publishing Bristol and Philadelphia, London 1993, p.21;
Efimov G. V., {\it On bound states in quantum field theory}, Bogoliubov
Laboratory of Theoretical Physics, JINR, 141980 Dubna, Russia,
hep--ph/9607425 25 July 1996, 33p.
\end{thebibliography}
\end{document}